\documentstyle[12pt,epsfig]{article}

\newlength{\dinwidth}
\newlength{\dinmargin}
\begin{document}
\def\bold#1{\setbox0=\hbox{$#1$}%
     \kern-.025em\copy0\kern-\wd0
     \kern.05em\%\baselineskip=18ptemptcopy0\kern-\wd0
     \kern-.025em\raise.0433em\box0 }
\def\slash#1{\setbox0=\hbox{$#1$}#1\hskip-\wd0\dimen0=5pt\advance
         to\wd0{\hss\sl/\/\hss}}
         \def \de {\partial}
\def \l {\lambda}
\def \L {\Lambda}
\def \T {\Theta}
\def \X {\Xi}
\def \D {\Delta}
\def \b {\beta}
\def \a {\alpha}
\def \G {\Gamma}
\def \g {\gamma}
\def \c {\chi}
\def \h {\eta}
\def \s {\sigma}
\def \S {\Sigma}
\def \d {\delta}
\def \D {\Delta}
\def \e {\varepsilon}
\def \ep {\epsilon}
\def \r {\rho}
\def \o {\omega}
\def \O {\Omega}
\def \t {\vartheta}
\def \te {\theta}
\def \ta {\tau}
\def \m {\mu}
\def \n {\nu}
\def \z {\zeta}
\def \f {\varphi}
\def \F {\Phi}
\def \x {\xi}
\def \non {\nonumber}
\def \noi {\noindent}
\def \ra {\rightarrow}
\def \pr {\prime}
\def \im {\Rightarrow}
\def \eq {\Leftrightarrow}
\def \mau {\geqslant}
\def \miu {\leqslant}
\def \dps {\displaystyle}
\def \fr {\displaystyle\frac}
\def \sh {\slashed}
\def\laq{~\raise 0.4ex\hbox{$<$}\kern -0.8em\lower 0.62 ex\hbox{$\sim$}~}
\def\gaq{~\raise 0.4ex\hbox{$>$}\kern -0.7em\lower 0.62 ex\hbox{$\sim$}~}
\def \lab {\label}
\def \fa {\forall}
\def \es {\exists}

\newcommand{\be}{\begin{equation}}
\newcommand{\ee}{\end{equation}}
\newcommand{\bea}{\begin{eqnarray}}
\newcommand{\eea}{\end{eqnarray}}
\newcommand{\nn}{\nonumber}
\newcommand{\dd}{\displaystyle}
\newcommand{\bra}[1]{\left\langle #1 \right|}
\newcommand{\ket}[1]{\left| #1 \right\rangle}
\newcommand{\spur}[1]{\not\! #1 \,}
\thispagestyle{empty} \vspace*{1cm}
\rightline{BARI-TH/538-06}\vspace*{2cm}
\begin{center}
  \begin{LARGE}
  \begin{bf}
$D_{sJ}(2860)$ resonance\\ \vspace*{0.4cm} and the $s_\ell^P={5\over 2}^-$  $c \bar s$ ($c \bar q$)  doublet
  \end{bf}
  \end{LARGE}
\end{center}
\vspace*{8mm}
\begin{center}
\begin{large}
P. Colangelo$^a$,  F. De Fazio$^a$ and  S. Nicotri$^{a,b}$
\end{large}
\end{center}
\begin{center}
\begin{it}
$^a$Istituto Nazionale di Fisica Nucleare, Sezione di Bari,
Italy\\ $^b$ Universit\'a degli Studi di Bari, Italy
\end{it}
\end{center}
\begin{quotation}
\vspace*{1.5cm}
\begin{center}
  \begin{bf}
  Abstract\\\end{bf}
\end{center}
\vspace*{0.5cm} \noindent
We analyze  the recently discovered  $D_{sJ}(2860)$ meson
using  arguments based on the heavy quark expansion together with the experimental observations.
We consider the  possible  quantum number  assignments,  in particular
 the interpretation of the meson as the $J^P=3^-$ $c \bar s$ state. 
\end{quotation}
\newpage
\baselineskip=18pt \vspace{2cm} \noindent

The observation of the meson  $D_{sJ}(2860)$, recently announced by the BaBar Collaboration \cite{palano06},   provides us with another interesting piece of information about  the $c \bar s$ system. The reported features of the new state are the following:
\begin{itemize}
\item  it  is observed in the $DK$ system inclusively produced in the process
$e^+ e^- \to DKX$; 
\item
it is  reconstructed in $D^0 K^+$, with $D^0$ both decaying to $K^- \pi^+$ and $K^- \pi^+ \pi^0$, and in $D^+ K_S^0$;
\item
 the resonance parameters are:
\bea
M(D_{sJ}(2860))&=&2856.6 \pm 1.5 \pm 5.0 \,\, MeV  \nn \\
\Gamma(D_{sJ}(2860) \to DK)&=& 47\pm 7 \pm10 \,\, MeV \,\,\, 
\eea
where $DK$ represents the sum of  the $D^0 K^+$ and $D^+ K_S^0$ modes.
\end{itemize}
Together with these features, the BaBar Collaboration noticed that:
\begin{itemize}
\item
no structures seem to appear in the $D^*K$ invariant mass distribution in the same range of mass
($\simeq 2860$ MeV);
\item
an additional broad contribution seems to be present  in the $DK$ 
distribution;  if fitted by a Breit-Wigner form,
it corresponds to the parameters $M=2688\pm4\pm3$ MeV and $\Gamma=112\pm7\pm36$ MeV.
\end{itemize}

The Collaboration also reported  new measurements of the  properties of the spin two
$D_{s2}(2573)$ meson, in particular of the  decay width 
$\Gamma(D_{s2}\to DK)$:
$\Gamma(D_{s2}(2573) \to DK)=27.1\pm0.6\pm5.6$ MeV. Considering the other possible
decay modes of $D_{s2}(2573)$, this value essentially corresponds to the meson full width; 
it must be compared to the previous value reported by the Particle Data Group:
$\Gamma(D_{s2}(2573))=15^{+5}_{-4}$ MeV  \cite{PDG}.

To interpret the new $c \bar s$ resonance we observe  that
the analysis of mesons comprising a single heavy quark $Q$ is  simplified in the heavy quark limit $m_Q \to \infty$. Indeed, in this
limit  the spin $s_Q$ of the heavy quark  and  the angular momentum
 $s_\ell$ of the meson light degrees of freedom: $s_\ell=s_{\bar q}+ \ell$ ($s_{\bar q}$ 
being the light antiquark spin and $\ell$ the orbital angular
momentum of the light degrees of freedom relative to the heavy quark) are decoupled in QCD, so that  the spin-parity  $s_\ell^P$ 
of the light degrees of freedom is conserved in strong interaction processes
\cite{HQET}.  Mesons can be classified as doublets of  $s_\ell^P$. Two states
correspond  to orbital angular momentum $\ell=0$;   in general  we  denote them  as
$(P,P^*)$ with $J^P=(0^-,1^-)$. The four states corresponding to  
$\ell=1$ can be collected in two  doublets, one  with  $ s_\ell^P={1 \over 2}^+$  and  $J^P=(0^+,1^+)$, and  another one with $ s_\ell^P={3 \over 2}^+$ and $J^P=(1^+,2^+)$; the states in the  two doublets are generically denoted as
$(P^*_{0},P_{1}^\prime)$ and $(P_{1},P_{2})$, respectively. For $\ell=2$ the doublets have $s_\ell^P={3 \over 2}^-$ and ${5 \over 2}^-$.

In case of charm, the heavy quark mass $m_c$ is greater than the strong interaction scale $\Lambda_{QCD}$ but it is not very large; therefore, 
corrections can be expected compared to the infinite limit. Two $O({1\over m_c})$ effects affecting the spectroscopy of the six $\ell=0,1$ states are the hyperfine splitting between mesons belonging to the same $s_\ell^P$ doublet,
and the  mixing of the two axial vector states  with $s_\ell^P{1 \over 2}^+$ and ${3 \over 2}^+$  to provide the two physical states.
 However, it is noticeable that the six  known  $c \bar s$ states reported by the Particle Data Group \cite{PDG}
can be classified according to this scheme as doublets of   $s_\ell^P$, as reported in Table \ref{doublets}.
\begin{table}[ht]
    \caption{$c \bar s$ states organized according to $s_\ell^P$ and $J^P$. The mass of  known mesons    is indicated. }
    \label{doublets}
    \begin{center}
    \begin{tabular}{c|cccccc}
      \hline 
 $s_\ell^P  $  &${1\over 2}^-$  &   ${1\over 2}^+$ &  ${3\over 2}^+$  & ${3\over 2}^-$ &  ${5\over 2}^-$ \\ \hline
 $J^P=s_\ell^P-{1\over 2}$&$D_s (1965) \,\, (0^-)  $  & $D^*_{sJ}(2317) \,\, (0^+)$ &  $D_{s1}(2536) \,\, (1^+) $ & $(P^{*\prime}_{s1})  \,\, (1^-)$& $(P^{*\prime}_{s2}) \,\,(2^-)$\\
 $J^P=s_\ell^P+{1\over 2}$&$D_s^*(2112) \,\, (1^-)$ & $D_{sJ}(2460)   \,\, (1^+)$  & $D_{s2}(2573) \,\,
  (2^+)  $ & $(P^{*}_{s2}) \,\, (2^-)$& $ (P_{s3}) \,\, (3^-)$\\
  \hline
    \end{tabular}
  \end{center}
\end{table}
As a matter of fact,  $D_s$  and  $D^*_s$ correspond to the  states belonging to
 the  $ s_\ell^P={1 \over 2}^-$  $c \bar s$ doublet.  For the four states  corresponding to  
$\ell=1$ we have   four  candidates:   $D^*_{sJ}(2317)$ ($J^P=0^+)$, 
$D_{sJ}(2460)$ and  $D_{s1}(2536)$ ($J^P=1^+$), and  $D_{s2}(2573)$  ($J^P=2^+$).  The natural assignment is 
$D^*_{sJ}(2317)$  to the  $ s_\ell^P={1 \over 2}^+$  $c \bar s$ doublet and $D_{s2}(2573)$ to the  $ s_\ell^P={3 \over 2}^+$  $c \bar s$ doublet
\footnote{We  consider $D^*_{sJ}(2317)$ and $D_{sJ}(2460)$ as  ordinary $c \bar s$ states. Other interpretations of this controversial resonances 
 are described in \cite{Colangelo:2003vg}.}. 
As for $D_{sJ}(2460)$ and  $D_{s1}(2536)$, they can be a mixing of the $1^+$ $ s_\ell^P={1 \over 2}^+$ and  ${3 \over 2}^+$ $c \bar s$
states. However, in the case of the non strange axial-vector $c \bar q$ mesons  the mixing angle has been measured and it is small:
$\omega=-0.10\pm 0.03\pm0.02\pm0.02$ rad \cite{mixing}, a result confirmed by an analysis of $O({1\over m_c})$ effects breaking the heavy quark spin symmetry
\cite{Colangelo:2005gb}.   Invoking $SU(3)_F$, also the mixing angle in the case of $c \bar s$ is expected to be small, so that 
the two $1^+$ states $D_{s1}(2536)$ and $D_{sJ}(2460)$  essentially coincide with  the   $s_\ell^P={3 \over 2}^+$ and
 ${1 \over 2}^+$ states.  \footnote{The mixing angle between the two $1^+$ states turns out to be sizeable in 
the unitarized meson model  in \cite{vanbeveren1}.} 
A puzzling aspect to be mentioned is that the two $s_\ell^P={1 \over 2}^+$ 
 $c \bar s$ states are almost degenerate in mass with the corresponding non-strange mesons,
 a physical effect not reproduced by the calculation of chiral corrections to the meson masses
 \cite{becirevic}.

In this classification a new $c \bar s$ meson decaying in two pseudoscalar mesons ($DK$) can be either  the
$J^P=1^-$ state of the  $ s_\ell^P={3 \over 2}^-$ doublet,  or the 
$J^P=3^-$ state of the  $ s_\ell^P={5 \over 2}^-$ doublet; in both cases it would correspond,
 in the constituent quark model,
to a state with orbital angular momentum $\ell=2$ and  lowest radial quantum number. As a matter of fact, 
we expect that these doublets  have lower mass than  all the  other  states with higher $s_\ell^P$. 
The other possibility is that $D_{sJ}(2860)$ is a radial excitation of already observed $c\bar s$ mesons:
 it could be the  $J^P=1^-$   $ s_\ell^P={1 \over 2}^-$ state (the first radial excitation of $D_s^*$),  a  $J^P=0^+$   $ s_\ell^P={1 \over 2}^+$ state (the first radial excitation of $D_{sJ}^*(2317)$) or
a   $J^P=2^+$   $ s_\ell^P={3 \over 2}^+$ state (the first radial excitation of $D_{s2}(2573)$). The analysis of the helicity distribution of the final state would be useful in constraining the various possibilities for the quantum numbers of the particle. In the absence of that,  arguments can be provided to  support a particular assignment of $J^P$  on the basis of the observed mass,  the decay modes and width.

\begin{figure}[t]
 \begin{center}
  \includegraphics[width=0.4\textwidth] {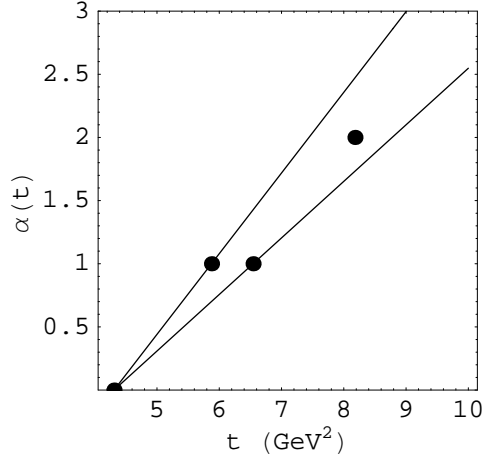}
\vspace*{0mm}
 \caption{Chew-Frautschi plot for the $c \bar s$ states. The point at $\alpha=0$ corresponds to the
 doublet $s_\ell^P={1\over 2}^-$ ($t=\overline{M}_H^2$), at $\alpha=1$ to the states  with $s_\ell^P={1\over 2}^+,{3\over 2}^+$ ($t=\overline{M}_{S,T}^2$). The point at $\alpha=2$ 
 corresponds to $D_{sJ}(2860)$.  }
  \label{regge}
 \end{center}
\end{figure}

A first information comes from  the  mass of the resonance. Let us consider the spin averaged mass
of the mesons belonging to the $ s_\ell^P={1 \over 2}^\mp$ and $ s_\ell^P={3 \over 2}^+$ doublets:
\be{\overline M}_{H}= {3 M_{P^*}+M_{P}  \over 4} \,\,\,, \,\,\,
 {\overline M}_{S} = {3 M_{P^*_0}+M_{P_1^\prime}  \over 4} \,\,\, , \,\,\,
{\overline M}_{T} = {5 M_{P_2}+3M_{P_1} \over 8}\,\,\, .
\label{lambdapar}\ee
Their numerical values  are 
${\overline M}_{H}= 2.0761(5)$ GeV, 
$ {\overline M}_{S} = 2.424(1)$ GeV and 
${\overline M}_{T} = 2.558(1)$ GeV. Using these values, we can draw
a Chew-Frautschi plot:  $ \alpha=l$ {\it vs}  $t$,  with a point at $(t,\alpha)=({\overline M}_{H}^2,0)$ 
and the points $(t,\alpha)=( {\overline M}_{S,T}^2,1)$. A Regge
trajectory $\alpha(t)=\alpha_0+\alpha^\prime t$ is obtained, with parameters:
$\alpha_0=-2.75, \,\,\, \alpha^\prime=0.64\,\,GeV^{-2}$ or
$\alpha_0=-1.93, \,\,\, \alpha^\prime=0.45\,\,GeV^{-2}$,  as shown in fig.\ref{regge}. Notice that, neglecting the light quark  spin-orbit interaction,  $ {\overline M}_{S}$ and ${\overline M}_{T}$
should be equal.
 %
%
If we attribute to $D_{sJ}(2860)$  the  orbital angular momentum $\ell=2$ we find  in the same plot  that the point $(t,\alpha)=({M}_{D_{sJ}(2860)}^2,2)$ belongs 
 to  a region comprised by the two lines,  an expected feature if  
the assignment is correct.  This observation does not  exclude that, being a radial excitation, the particle belongs to a different trajectory, but it can be used as a first hint to support the assignment to the $\ell=2$ 
 $c \bar s$  states.

Another piece of information comes from the measured $DK$ width. In ref. \cite{Colangelo:2000jq}
it was  suggested that a few high mass and high spin charm states  could be narrow
enough to be observed. In particular, it was suggested that the $3^-$ state  belonging to the
$s_\ell^P={5\over 2}^-$  $c \bar q$ ($c \bar s$) doublet  should be not too broad  since it decays to $D \pi$ ($D K$) in $f-$wave. 

We can
elaborate on this point,  using an analysis based on the heavy
quark limit  \cite{positivep}.  We define
 the fields representing the various heavy-light  meson doublets:    
$H_a$ for $s_\ell^P={1\over2}^-$ ($a=u,d,s$ is a light flavour index),
$S_a$ and $T_a$ for $s_\ell^P={1\over2}^+$ and $s_\ell^P={3\over2}^+$, respectively;  we also
consider the fields $X_a$ and $X^\prime_a$ for the doublets  corresponding  to orbital angular momentum $\ell=2$, 
i.e.  $s_\ell^P={3\over2}^-$ (we denote the states $(1^-,2^-)$ as $(P^{*\prime}_1,P^*_2)$) and $s_\ell^P={5\over2}^-$ (the states $(2^-,3^-)$ denoted as $(P^{*\prime}_2,P_3)$), respectively:
\bea
H_a & =& \frac{1+{\rlap{v}/}}{2}[P_{a\mu}^*\gamma^\mu-P_a\gamma_5]  \label{neg} \nn  \\
S_a &=& \frac{1+{\rlap{v}/}}{2} \left[P_{1a}^{\prime \mu}\gamma_\mu\gamma_5-P_{0a}^*\right]   \nn \\
T_a^\mu &=&\frac{1+{\rlap{v}/}}{2} \left\{ P^{\mu\nu}_{2a} \gamma_\nu -P_{1a\nu} \sqrt{3 \over 2} \gamma_5 \left[
g^{\mu \nu}-{1 \over 3} \gamma^\nu (\gamma^\mu-v^\mu) \right]
\right\}  \label{pos2} \\
X_a^\mu &=&\frac{1+{\rlap{v}/}}{2} \left\{ P^{*\mu\nu}_{2a} \gamma_5 \gamma_\nu -P^{*\prime}_{1a\nu} \sqrt{3 \over 2}  \left[
g^{\mu \nu}-{1 \over 3} \gamma^\nu (\gamma^\mu-v^\mu) \right]
\right\}   \nn   \\
X_a^{\prime \mu\nu} &=&\frac{1+{\rlap{v}/}}{2} \left\{ P^{\mu\nu\sigma}_{3a} \gamma_\sigma -P^{*'\alpha\beta}_{2a} \sqrt{5 \over 3} \gamma_5 \left[
g^\mu_\alpha g^\nu_\beta -{1 \over 5} \gamma_\alpha g^\nu_\beta (\gamma^\mu-v^\mu)-  {1 \over 5} \gamma_\beta g^\mu_\alpha (\gamma^\nu-v^\nu) \right]
\right\}  \nn
\eea
with the various operators annihilating mesons of four-velocity $v$
which is conserved in  strong interaction processes (the heavy field operators  contain a
factor $\sqrt{m_P}$ and have dimension $3/2$). The interaction of these particles with the
octet of light pseudoscalar mesons, introduced using 
 $\displaystyle \xi=e^{i {\cal M} \over
f_\pi}$, $\Sigma=\xi^2$ and  the matrix ${\cal M}$ containing
$\pi, K$ and $\eta$ fields:
\begin{equation}
{\cal M}= \left(\begin{array}{ccc}
\sqrt{\frac{1}{2}}\pi^0+\sqrt{\frac{1}{6}}\eta & \pi^+ & K^+\nonumber\\
\pi^- & -\sqrt{\frac{1}{2}}\pi^0+\sqrt{\frac{1}{6}}\eta & K^0\\
K^- & {\bar K}^0 &-\sqrt{\frac{2}{3}}\eta
\end{array}\right)
\end{equation}
($f_{\pi}=132 \; $ MeV) can be described by an effective Lagrangian 
which is invariant under chiral and
 heavy-quark spin-flavour transformations. The kinetic term of the Lagrangian
 \begin{eqnarray} {\cal L} &=& i\; Tr\{ {\bar H}_b v^\mu
D_{\mu ba}  H_a \}  + \frac{f_\pi^2}{8}
Tr\{\partial^\mu\Sigma\partial_\mu \Sigma^\dagger \} \nn \\ &+&
\sum_{F=S,T^\mu,X^\mu,X^{\prime \mu \nu}} Tr\{ {\bar F}_b \;( i \; v^\sigma D_{\sigma ba} \; - \; \delta_{ba} \;
\Delta_F)  F_a \}
  \label{L}
\end{eqnarray}
involves the operators $D$ and $\cal A$:
\begin{eqnarray}
D_{\mu ba}&=&-\delta_{ba}\partial_\mu+{\cal V}_{\mu ba}
=-\delta_{ba}\partial_\mu+\frac{1}{2}\left(\xi^\dagger\partial_\mu
\xi
+\xi\partial_\mu \xi^\dagger\right)_{ba}\\
{\cal A}_{\mu ba}&=&\frac{i}{2}\left(\xi^\dagger\partial_\mu
\xi-\xi
\partial_\mu \xi^\dagger\right)_{ba} 
\end{eqnarray}
and the mass parameters $\Delta_F$ which represent
the mass splittings between   the excited  and the low-lying 
(negative parity) doublets expressed
 in terms of the spin-averaged masses:
 $\Delta_F= \overline M_F - \overline M_H$.  For $H, S$ and $T$ doublets 
 the spin-averaged masses $\overline M_{H,S,T}$ are given in
 eq.(\ref{lambdapar}), while for the doublets $X$ and $X^\prime$  they read:
 \be
 {\overline M}_{X} = {5 M_{P^{*\prime}_2}+3M_{P^{*\prime}_1} \over 8} \,\,\,\, , \,\,\,\,
{\overline M}_{X^\prime} ={7 M_{P_3}+ 5 M_{P^*_2} \over 12} \,\, .  
\label{lambdapar1}\ee

At the leading order in the heavy quark mass and light meson momentum expansion
the decays  $F \to H M$ $(F=H,S,T,X,X^\prime$ and $M$ a light pseudoscalar meson) can be described by the Lagrangian interaction  terms \cite{hqet_chir}: 
\bea
{\cal L}_H &=& \,  g \, Tr [{\bar H}_a H_b \gamma_\mu \gamma_5 {\cal
A}_{ba}^\mu ] \nn \\
{\cal L}_S &=& \,  h \, Tr [{\bar H}_a S_b \gamma_\mu \gamma_5 {\cal
A}_{ba}^\mu ]\, + \, h.c. \,, \nn \\
{\cal L}_T &=&  {h^\prime \over \Lambda_\chi}Tr[{\bar H}_a T^\mu_b
(i D_\mu {\spur {\cal A}}+i{\spur D} { \cal A}_\mu)_{ba} \gamma_5
] + h.c.   \label{lag-hprimo}   \\
{\cal L}_X &=&  {k^\prime \over \Lambda_\chi}Tr[{\bar H}_a X^\mu_b
(i D_\mu {\spur {\cal A}}+i{\spur D} { \cal A}_\mu)_{ba} \gamma_5
] + h.c.  \nn \\
{\cal L}_{X^\prime} &=&  {1 \over {\Lambda_\chi}^2}Tr[{\bar H}_a X^{\prime \mu \nu}_b
[k_1 \{D_\mu, D_\nu\} {\cal A}_\lambda+ k_2 (D_\mu D_\nu { \cal A}_\lambda + D_\nu D_\lambda { \cal A}_\mu)]_{ba}  \gamma^\lambda \gamma_5] + h.c.  \nn
\eea
where $\Lambda_\chi$ is  the chiral symmetry-breaking scale; we  use
$\Lambda_\chi = 1 \, $ GeV.
${\cal L}_S$ and ${\cal L}_T$ describe transitions of positive parity heavy mesons with
the emission of light pseudoscalar mesons in $s-$ and $d-$ wave, respectively, $g, h$ and $h^\prime$
representing  effective coupling constants.  On the other hand,
${\cal L}_X$ and ${\cal L}_{X^\prime}$  describe the
transitions of higher mass mesons of negative parity with
the emission of light pseudoscalar mesons in $p-$ and $f-$ wave
with  coupling constants $k^\prime$, $k_1$ and $k_2$. We only consider these terms:
the light meson momenta involved in the $D_{sJ}(2860)$ decays are $q_K=0.59 $ GeV for $D^*K$ final state
  and $q_K=0.7 $ GeV for $DK$ final state, so  it is possible that other terms in the light-meson momentum expansion, involving other structures and couplings,  should be taken into account
in the interaction Lagrangian. However, at present these terms are  unknown,
therefore we only consider the interaction terms in  eq.(\ref{lag-hprimo}).

At the same  order in the expansion in the light meson momentum,
the structure of the Lagrangian terms for radial excitations of the $H$, $S$ and $T$ doublets does not
change,  since it  is only dictated by the spin-flavour and chiral symmetries,  
but  the coupling constants $g, h$ and $h^\prime$ have to be  substituted by $\tilde g, \tilde h$ and $\tilde h^\prime$.  
The advantage of this formulation is that meson transitions  into final states obtained  by $SU(3)$ and heavy quark spin rotations can be related in a straightforward way.

In Table \ref{ratios} we report  the ratios 
$\displaystyle {\Gamma( D_{sJ} (2860) \to D^*K) \over \Gamma( D_{sJ} (2860)\to DK) }$ 
and  $\displaystyle {\Gamma( D_{sJ} (2860)\to D_s \eta) \over \Gamma( D_{sJ} (2860)\to DK)  }$
obtained for various     quantum number assignments to $D_{sJ} (2860)$  using
 eqs.(\ref{pos2}) and (\ref{lag-hprimo}). The ratios
 do not depend on the coupling constants,  but only on the quantum numbers
and on the kinematics of the various processes. 
\footnote{In our framework, calculations of widths do not include possible effects  such as couplings to meson-meson continuum. An  approach to these effects is described in a
model  in  \cite{vanbeveren1} and  references therein.} 
\begin{table}[ht]
    \caption{Predicted ratios
    $\displaystyle {\Gamma( D_{sJ} \to D^*K) \over \Gamma( D_{sJ} \to DK)
 }$  and   $\displaystyle {\Gamma( D_{sJ} \to D_s \eta) \over \Gamma( D_{sJ} \to DK)  }$ for the various assignment
 of quantum numbers to  $D_{sJ}(2860) $.  For $DK$ we mean the sum $DK=D^0K^+ + D^+ K_S^0$.}
    \label{ratios}
    \begin{center}
    \begin{tabular}{| c | c | c | c |}
      \hline
 $D_{sJ}(2860) $  &  $D_{sJ}(2860) \to DK $&$\displaystyle{\Gamma( D_{sJ} \to D^*K) \over \Gamma( D_{sJ} \to DK)
 }$  &   $\displaystyle{\Gamma( D_{sJ} \to D_s \eta) \over \Gamma( D_{sJ} \to DK)  }$ 
\\ \hline
 $s_\ell^p={1\over 2}^-$, $J^P=1^-$,  $n=1$ & $p$-wave &$1.23$& $0.27$ \\
$s_\ell^p={1\over 2}^+$, $J^P=0^+$, $n=1$&  $s$-wave &$0$& $0.34$ \\
$s_\ell^p={3\over 2}^+$, $J^P=2^+$, $n=1$&  $d$-wave &$0.63$& $0.19$\\
$s_\ell^p={3\over 2}^-$, $J^P=1^-$,   $n=0$ & $p$-wave  & $0.06$& $0.23$ \\
$s_\ell^p={5\over 2}^-$, $J^P=3^-$,   $n=0$ & $f$-wave  & $0.39$& $0.13$ \\
    \hline
    \end{tabular}
  \end{center}
\end{table}

The ratios   in Table  \ref{ratios} must be considered together with the experimental observation. 
In particular, non observation (at present)  of a $D^*K$ signal in the $D_{sJ}$ range of mass implies that
the production of $D^* K$ is not favoured, and therefore  the assignments
$s_\ell^p={1\over 2}^-$, $J^P=1^-$,  $n=1$,  and $s_\ell^p={3\over 2}^+$, $J^P=2^+$, $n=1$
can be excluded:
  $D_{sJ}$ is not a radial excitation of $D_s^*$ and of $D_{s2}$. 

The assignment 
$s_\ell^p={3\over 2}^-$, $J^P=1^-$, $n=0$ can also be excluded, even though in this case the decay
into $D^*K$ would be suppressed.  As a matter of fact,  the  width  $\displaystyle
 \Gamma(D_{sJ}\to DK)= {8 \over 3} {k^{\prime 2} \over \pi \Lambda_\chi^2 f_\pi^2} {M_D \over M_{D_{sJ}}} E_K^2 q_K^3 \,\,\, $ ($E_K$ and $q_K$ being kaon energy and momentum in the $D_{sJ}$
 rest frame),
 obtained using the relevant  term of the interaction Lagrangian in (\ref{lag-hprimo}), would be  $\Gamma(D_{sJ}\to DK)\simeq 1.5$ GeV if we use
a coupling $k^\prime \simeq h^\prime\simeq 0.45\pm0.05$,  as  the $h^\prime$   was determined in \cite{Colangelo:2005gb}. The large width is expected  since the transition $D_{sJ} \to DK $ occurs in
this case in $p-$wave.
There is no  reason  to presume that the coupling constant $k^\prime$ is sensibly smaller,
 so that
the small value of  $\Gamma(D_{sJ}\to DK)$ seems 
incompatible with a  $p-$wave transition.  

In the case of the assignment  $s_\ell^p={1\over 2}^+$, $J^P=0^+$, $n=1$ 
the decay  $D_{sJ}\to D^*K$ is forbidden. On the other hand,  $D_{sJ}\to DK$ would occur 
in $s-$wave, with
$\displaystyle
\Gamma(D_{sJ}\to DK)= {3 \over 4} {\tilde h^{2} \over \pi f_\pi^2} {M_D \over M_{D_{sJ}}} E_K^2 q_K \,\,\, $.
For the state with the lowest radial quantum
number $n=0$ the coupling costant was computed:  $h\simeq-0.55$ with
phenomenological consequences  essentially in agreement  with observation 
\cite{Colangelo:1995ph,Colangelo:2003vg}; using this  value for $\tilde h$ we would obtain  $\Gamma(D_{sJ}\to DK)\simeq 1.4$ GeV.  But it is reasonable to suppose that  $|\tilde h| < |h|$,  although no information is  available about the couplings of radially excited heavy-light mesons to low-lying states. Using the experimental width we would obtain $\tilde h=0.1$.   A large signal would  be expected in the $D_s \eta$ channel, as reported in Table   \ref{ratios}.  
A problem with this assignment is the following. If $D_{sJ}(2860)$ is a scalar radial excitation, it should have  a spin partner with 
 $J^P=1^+$ ($s_\ell^p={1\over 2}^+$, $n=1$)  decaying to $D^* K$ with a small width, of the order of 40 MeV, a rather easy signal to detect.
 For $n=0$ both the states $D^*_{sJ}(2317)$ and $D_{sJ}(2460)$ are produced in charm continuum  at  $e^+ e^-$ factories. 
To explain the absence of the $D^*K$ in charm continuum events  at mass around $2860$ MeV,
one should  invoke some mechanism favouring the production of the $0^+$ $n=1$ state and inhibiting the production of $1^+$ $n=1$ state,   
 a mechanism which discriminates the first radial excitation from the low lying state $n=0$. We consider a mechanism of this kind  difficult to imagine.
Finally, this assignment  seems not compatible with  the argument concerning the meson mass provided above. 
\footnote{The  interpretation of
$D_{sJ}(2860)$ as the first radial excitation of $D^*_{sJ}(2317)$  has been proposed 
in  \cite{vanBeveren:2006st} within a  unitarized meson model, and in \cite{Close:2006gr}
in a quark model with modified spin-orbit and spin-spin interaction.}
   
Let us consider the last possibility:  $s_\ell^p={5\over 2}^-$, $J^P=3^-$, $n=0$.
In this case, the small $DK$ width is due to the huge suppression related to the kaon momentum factor:
$\displaystyle
\Gamma(D_{sJ}\to DK)= {6 \over 35} {(k_1+k_2)^{2} \over \pi f_\pi^2 \Lambda_\chi^4} {M_D \over M_{D_{sJ}}}  q_K^7$.  A smaller but non negligible signal
in the $D^*K$ mode is predicted,  and  a small signal in the $D_s \eta$ mode is also expected,
see Table \ref{ratios}. From the observed width, the combination of the coupling constants $k_1 + k_2$ 
turns out to be $k_1+k_2=0.52$, similar to the  couplings of the other doublets to light
pseudoscalars.
 For the same value of the coupling, the state of spin two  belonging to  the $s_\ell^P={5 \over 2}^-$ doublet, 
$D_{s2}^{*\prime}$, which has  $J^P=2^-$ and can  decay to $D^* K$ and not to $DK$,
would  be narrow: $\Gamma(D_{s2}^{*\prime}\to D^*K)\simeq 50$ MeV. But this is only true in the 
$m_Q\to \infty$ limit, where the $J^P=2^-$ state transition into $D^* K$ occurs in $f$-wave.
As an effect of $1/m_Q$ corrections the $D_{s2}^{*\prime}\to D^*K$ decay can occur in $p$-wave, 
in which case  the  width of  $D_{s2}^{*\prime}$ could be  broader;  therefore,  it is not necessary
to invoke a mechanism inhibiting the production of this state with respect to $J^P=3^-$.

If  $D_{sJ}(2860)$ has $J^P=3^-$, it is not expected to  be produced
 in non leptonic $B$ decays such as 
$B^0 \to  D^- D_{sJ}(2860)^+$ and $B^+ \to  \bar D^0 D_{sJ}(2860)^+$:    the
non leptonic amplitude in the factorization approximation vanishes   since
the vacuum matrix element of the weak $V-A$ current with a spin three particle is zero.
Therefore,  the quantum number assignment can be confirmed by studies of $D_{sJ}$ production in $B$
transitions. 
\footnote{After completion of this work,  Belle Collaboration  \cite{unknown:2006xm} has reported the evidence of a rather broad  state, $D_{sJ}(2715)$,
with $M=2715\pm11^{+11}_{-14}$ MeV,  $\Gamma=115\pm20^{+36}_{-32}$ MeV and $J^P=1^-$,
 found in the 
Dalitz plot analysis of $B^+ \to \bar D^0 D^0 K^+$. It is worth noticing that no signal of $D_{sJ}(2860)$ is found in the analysis of this decay mode.}

The conclusion  of our study  is that $D_{sJ}(2860)$ is likely  a $J^P=3^-$ state,
 a  predicted high mass, high spin and relatively narrow $c \bar s$ state   \cite{Colangelo:2000jq}. Its non-strange partner  $D_3$, if the mass splitting 
$M_{D_{sJ}(2860)}-M_{D_3}$ is of the order of the strange quark mass, is  also expected to be narrow:
$\Gamma(D_{3}^+\to D^0 \pi^+ )\simeq 37$ MeV. It  can  be 
produced in semileptonic as well as in non leptonic $B$ decays,
such as $B^0 \to D_3^- \ell^+  \bar \nu_\ell$ and $B^0 \to D_3^- \pi^+$
\cite{Colangelo:2000jq}: its observation could  be used to confirm the quantum number assignment to the  resonance $D_{sJ}(2860)$ found by BaBar.

\vspace*{1cm}

\noi{\bf Acknowledgments} We thank A. Palano and T.N. Pham for discussions.
One of us (PC) thanks CPhT,  \'Ecole Polytechnique,  for kind hospitality during the completion of this work. We acknowledge partial support from the EC Contract No.
HPRN-CT-2002-00311 (EURIDICE).

\newpage


\begin{thebibliography}{99}

\bibitem{palano06}
 B.~Aubert  [BABAR Collaboration],
  arXiv:hep-ex/0607082.
  

\bibitem{PDG}
  W.~M.~Yao {\it et al.}  [Particle Data Group],
  J.\ Phys.\ G {\bf 33} (2006) 1.
 
  \bibitem{HQET}
For reviews see: M.~Neubert,
  Phys.\ Rept.\  {\bf 245} (1994) 259;
  A.~V.~Manohar and M.~B.~Wise,
  Camb.\ Monogr.\ Part.\ Phys.\ Nucl.\ Phys.\ Cosmol.\  {\bf 10} (2000) 1;
  F.~De Fazio,
  in "At the frontier of Particle Physics/Handbook of QCD", edited by M. A. Shifman, World Scientific, Singapore,
  2001, p.1671 (arXiv:hep-ph/0010007).
  
  \bibitem{Colangelo:2003vg}
 P.~Colangelo, F.~De Fazio and R.~Ferrandes,
  Mod.\ Phys.\ Lett.\ A {\bf 19} (2004) 2083; 
E.~S.~Swanson,
Phys.\ Rept.\  {\bf 429} (2006) 243.

\bibitem{mixing}
  K.~Abe {\it et al.}  [Belle Collaboration],
  Phys.\ Rev.\ D {\bf 69} (2004) 112002.
  
\bibitem{Colangelo:2005gb}
  P.~Colangelo, F.~De Fazio and R.~Ferrandes,
  Phys.\ Lett.\ B {\bf 634} (2006) 235.

\bibitem{vanbeveren1}
 E.~van Beveren and G.~Rupp,
  Eur.\ Phys.\ J.\ C {\bf 32}  (2004) 493.

\bibitem{becirevic}
  D.~Becirevic, S.~Fajfer and S.~Prelovsek,
  Phys.\ Lett.\ B {\bf 599} (2004) 55.

\bibitem{Colangelo:2000jq}
  P.~Colangelo, F.~De Fazio and G.~Nardulli,
  Phys.\ Lett.\ B {\bf 478} (2000) 408.
  
 \bibitem{positivep}
N.Isgur and M.B.Wise, Phys. Rev. Lett. {\bf 66} (1991) 1130; Phys. Rev. {\bf D 43} (1991) 819;
U.Kilian, J.G.K\"orner and D.Pirjol, Phys. Lett. {\bf B 288} (1992) 360;
A.F.Falk and M.Luke, Phys. Lett. {\bf B 292} (1992) 119.

\bibitem{hqet_chir}
M.B.Wise, Phys. Rev. {\bf D 45}  (1992)  R2188;
G.Burdman and J.F.Donoghue, Phys. Lett. {\bf B 280} (1992) 287;
P.Cho, Phys. Lett. {\bf B 285} (1992)  145;
H.-Y.Cheng {\it et al.,}  Phys. Rev. {\bf D 46} (1992)  1148;
R.Casalbuoni {\it et al.,} Phys. Lett. {\bf B 299} (1993) 139.


\bibitem{Colangelo:1995ph}
  P.~Colangelo, F.~De Fazio, G.~Nardulli, N.~Di Bartolomeo and R.~Gatto,
  Phys.\ Rev.\ D {\bf 52} (1995) 6422;
P.~Colangelo and F.~De Fazio,
  Eur.\ Phys.\ J.\ C {\bf 4} (1998) 503;
P.~Colangelo and F.~De Fazio,
  Phys.\ Lett.\ B {\bf 570} (2003) 180.

\bibitem{vanBeveren:2006st}
  E.~van Beveren and G.~Rupp,
  arXiv:hep-ph/0606110.
  
\bibitem{Close:2006gr}
  F.~E.~Close, C.~E.~Thomas, O.~Lakhina and E.~S.~Swanson,
  arXiv:hep-ph/0608139.
  
\bibitem{unknown:2006xm}
 K. Abe {\it et al.}   [Belle Collaboration],
  arXiv:hep-ex/0608031.

\end{thebibliography}
\end{document}